# The CNT/BCN/CNT structure (zigzag type) as a molecular switch

H Milani Moghaddam

Department of Physics, University of Mazandaran, Babolsar, Iran

E-mail: milani@umz.ac.ir ; hossainmilani@yahoo.com

**Abstract.** Using a tight-binding model and some well-known approaches and methods based on Green's function theory and *Landauer* formalism, we numerically investigate the conductance properties and I-V characteristics of ($n$,0) zigzag single-walled BCN alloy nanotube in the CNT/BCN/CNT structure, where nanocontacts are considered as ($n$,0) zigzag single-walled carbon nanotubes. Our calculations show that any increasing in $n$ considerably give rise to the enhancing of the conductance of the system. With our system characteristics, this system can be a possible candidate for a nanoelectronic switching device.

**PACS Nos:** 73.63.Fg; 73.22.-f; 73.63.-b

**Keywords:** molecular wire, carbon nanotube, boron carbonitride alloy nanotube, Green's function

## 1. Introduction

The study and manipulation of matter on the nanometer scale is a thriving area of research, with profound implications for technology (e.g. nanoelectronics, nanostructured materials, nanobiology). So devices have been designed in such way that a molecule is sandwiched between two electrodes (metallic or organic).

So far, many theoretical works have been developed and models have been proposed to represent the molecular system and the reservoirs. In most model systems, the molecular wires connected to two semi-infinite surfaces [1, 2] or to two semi-infinite 'rods' [3] or to clusters [4, 5]. In some recent works, simple molecular wires have been connected to two semi-infinite carbon nanotubes (CNTs) [6-9]. The significantly improved switching characteristics of short organic FETs with metallic CNT electrodes over those with metal electrodes are attributed to the excellent electrostatics attainable with a nanotube electrode geometry [10, 11].

CNTs and hetero-materials including borons (B) and nitrogens (N) have been attracting much attention both in the fundamental science and in the interests of application to nanotechnology devices [12, 13, 14]. Though CNTs have generated great interest in use of a broad range of potential nanodevices for their unique structural and electronic properties [15, 16], other nanotubes such as boron carbonitride (BCN) alloy nanotubes are interesting in their own right and may be able to offer different possibilities for technological applications that CNTs cannot provide [17]. It is calculated that heterojunctions of B–C–N nanotubes are largely independent of the radius, helicity, multiplicity, or degree of perfection of the constituting nanotubes [17], though depending on their chirality, CNTs can be metallic or semiconducting [18].

BCN alloy nanotubes have been successfully synthesized by electrical arc discharge [19-23], pyrolysis [24, 25] and laser ablation [26] methods. Of all these properties, BCN alloy nanotubes are of especially importance for nanodevice applications.

In this letter, following these interests and towards the modeling of a molecular wire, we use a model in which a zigzag single-walled BCN alloy nanotube is connected to two semi-infinite zigzag single-walled CNTs which are considered as nanocontacts. In this model of BCN alloy nanotube, B atoms are replaced by C atoms in the middle unit cell of the BN nanotube. We numerically investigate the conductance properties and I-V characteristics of ($n$,0) BCN alloy nanotube in the CNT/BCN/CNT structure. The ($n$,0) zigzag CNTs (on which we will concentrate here) with $n = 3\ell$ ($\ell$ integer) will be metallic. The model and description of the methods for investigation of conductance properties of the molecular wire is introduced in Sec. (2). The results and discussion are presented in Sec. (3) followed by a summary in Sec. (4).

## 2. Methodology

The most commonly used computational schemes for calculating the (coherent) conductance g are the *Landauer* theory [27] and the Green's function formalism [28-30].

The conductance *g* at zero temperature is simply proportional to the transmission coefficient, $T(E)$, for injected electrons at the Fermi energy,

$$g = g_0 T(E) \; ; \; g_0 = \frac{2e^2}{h}. \tag{1}$$

We model the transport problem by dividing the system in three parts (Fig. 1): Two semi-infinite leads (L) and (R) with bulk electronic structure are connected to a finite region called device (D). the D region contains the scattering region (S) where the potential landscape for the electrons deviates from that in the leads, and a finite part of each lead (-1) and (1). The leads' parts inside D are chosen sufficiently big such that the leads only couple to that part of the device. The Hamiltonian matrix is divided into submatrices as follows

$$H = \begin{pmatrix} H_L & H_{LD} & 0 \\ H_{DL} & H_D & H_{DR} \\ 0 & H_{RD} & H_R \end{pmatrix}, \tag{2}$$

where it has been assumed that the electrodes do not directly interact which each other ($H_{LR} = 0 = H_{RL}$) as the D region is sufficiently big, or rather the electrode parts inside D as mentioned above .

The Green's function matrix is given by

$$G = \begin{pmatrix} G_L & G_{LD} & G_{LR} \\ G_{DL} & G_D & G_{DR} \\ G_{RL} & G_{RD} & G_R \end{pmatrix}. \tag{3}$$

We could calculate the above submatrices of the Green's function by solving the equation

$$(E - H) G(E) = 1. \tag{4}$$

This results in the following expression for the Green's function in the device region

$$G_D(E) = (E - H_D - \Sigma_L(E) - \Sigma_R(E))^{-1}. \tag{5}$$

The complex self-energies $\Sigma_L(E)$ and $\Sigma_R(E)$ describe the effect of the two leads on the electronic structure of the device and are given by the Green's functions of the semi-infinite isolated leads $g_L(E) = (E - H_L)^{-1}$ and $g_R(E) = (E - H_R)^{-1}$ projected into the device region by the coupling of the leads to the device $H_{DL}$ and $H_{RD}$

$$\Sigma_L(E) = H_{DL} g_L(E) H_{LD} \quad \text{and} \quad \Sigma_R(E) = H_{DR} g_R(E) H_{RD}. \tag{6}$$

It can be shown that the *Landauer* transmission at a certain energy can be expressed in the Green's function formalism by the Caroli expression [31]

$$T(E) = Tr[\Gamma_L(E)G_D^+(E)\Gamma_R(E)G_D(E)]. \tag{7}$$

The coupling matrices $\Gamma_L(E)$ and $\Gamma_R(E)$ are minus the imaginary part of the leads' self-energy.

$$\Gamma_{L(R)}(E) = i(\Sigma_{L(R)}(E) - \Sigma_{L(R)}^+(E)). \tag{8}$$

Also the electronic density of states (DOS) of the device is given as [31]

$$DOS(E) = -\frac{1}{\pi} \operatorname{Im}\{Tr[G_D(E)]\} \tag{9}$$

The CNTs and BCN alloy nanotube are modeled within the tight-binding Hamiltonian with only one $\pi$-orbital per atom [32, 33]. This Hamiltonian can describe reasonably well the band structure of a nanotube especially near the Fermi level, $\varepsilon_F$, which is zero in this case

$$H = \sum_j \varepsilon_j c_j^+ c_j - \sum_j t_{j+1,j}(c_{j+1}^+ c_j + c_j^+ c_{j+1}), \tag{10}$$

where $c_j(c_j^+)$ is the annihilation(creation) operator of an electron at the $j$ site. $\varepsilon_j$ and $t_{j+1,j}$, respectively, represent the on-site energy and the nearest-neighbor hopping integral.

## 3. Results and discussion

Based on the formalism described in section 2, we have investigated the electronic conduction properties of CNT/BCN/CNT structure for several typical CNT and BCN alloy nanotube with different features. In our calculations, the onsite energy at B atoms and that of N atoms are assumed to be $+2.33\,eV$ and $-2.50\,eV$, respectively, if being measured from the C onsite energy ($\varepsilon_C = 0$). Also we shall assume $t_{C-C} = -3\,eV$, $t_{B-C} = -2.7\,eV$ and $t_{B-N} = -2.81\,eV$ ($t$ stands for the hopping integral).

Fig. 2 illustrates the electronic density of states (DOS) of CNT/BCN/CNT system. The plots show that the presence of C atoms in BCN alloy nanotube induces the electronics states within the band gap and causes a large enhancement in the conductance of the system. Pure BN nanotubes are wide band gap semiconductors with a band gap of nearly $5\,eV$. Also the electronic transmission probability through the system is shown in Fig. 3, panels (a)-(c) which correspond to Fig. 2, panels (a)-(c), respectively. Our results suggest that any increasing in $n$ considerably give rise to the enhancing of the conductance of the system. From the most experimentally observed carbon nanotube sizes, there is a tiny gap in the zigzag nanotube types which arises from curvature effects [34]. Thus any increasing in the tube diameter give rises to decrease of the gap as $1/R^2$ [34,35] and causes a large enhancement in the conductance of the system.

In all of above calculations no voltage drop was considered across the system. However, in order to study the behavior of the system in the presence of an applied voltage we use

$$I(V) = (2e/h)\int_{\mu_L}^{\mu_R} d\varepsilon\, T(\varepsilon)[f_L(\varepsilon) - f_R(\varepsilon)] \tag{11}$$

where $I(V)$, $T(\varepsilon)$, $f_{L/R}(\varepsilon) = (\exp[\beta(\varepsilon - \mu_{L/R})]+1)^{-1}$ and $\mu_{L/R} = \varepsilon_F \pm eV/2$, respectively, represent the total current through the device in the V bias voltage, the transmission coefficient, the Fermi-Dirac distribution function and the chemical potential. $\beta$ is equal to $1/k_B T$ ($k_B$ and $T = 4°K$ are Boltzmann constant and temperature of the reservoir responsible for injecting the electrons into the contacts, respectively). $e$ and $h$ correspond to the electron's charge and Planck's constant in that order.

The results of these studies are presented in Fig. 4. The positions of the current jumps are determined by the position of the reservoir Fermi levels and by the details of the molecular structure of the wire. The appearance of jumps in the molecular Eigenvalue staircase is a conspicuous feature of the molecular wire I-V curve.

With our system characteristics, one can construct a switching device. It is clear immediately from Fig. 4 that by changing the voltage (e.g. by nearly 0.49 Volt) the device acts as a switch, and is turned from "OFF" to "ON". According to Fig. 4, the first jump ($V \approx 0.49\,Volt$) in the I-V characteristics is independent of the nanotubes' diameter but the current in that as the tube radius(R) increases, considerably increases (Fig. 5).

We explain the origin of the staircase shape of the I-V curve as follows. For small V, there are no molecular resonances between the Fermi levels of the two electrodes, and the current is small. As V increases, the energies of the wire orbitals decrease and eventually one of the molecular resonances crosses one of the Fermi levels. This opens a current channel, and shows as a jump in the I–V curve. This behavior is similar to the phenomenon of resonant tunneling observed in quantum well devices. One important difference, however, is the fact that in quantum wells, application of a voltage above the resonant voltage shuts down the current channel because the resonance lies in the energy gap of the reservoirs. With molecular wires, once a channel is activated, it remains open since the reservoirs are metallic and occupied states can always be found that align with the molecular resonance.

## 3. Summary

In brief, we numerically investigate the conductance properties and I-V characteristics of ($n$,0) zigzag single-walled BCN alloy nanotube in the CNT/BCN/CNT structure, where the nanocontacts are considered as ($n$,0) zigzag single-walled carbon nanotubes. We have applied some well-known approaches and methods based on Green's function theory and *Landauer* formalism as well as Tight-binding Hamiltonian model to investigate the electron conduction through the CNT/BCN/CNT structure.

Our results show that the presence of C atoms in BCN alloy nanotube induces the electronics states within the band gap and causes a large enhancement in the conductance of the system. Our outcomes suggest that any increasing in the nanotubes' diameter considerably give rise to the enhancing of the conductance of the system. Besides, according to our results, this device can be a possible candidate for a nanoelectronics switching device.

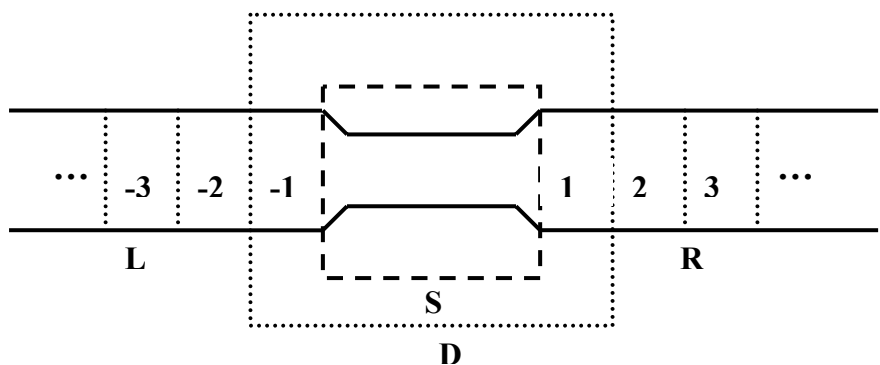

Fig. 1. A schematic representation of division of system into leads L and R, device D, and scattering region S as described in the text.

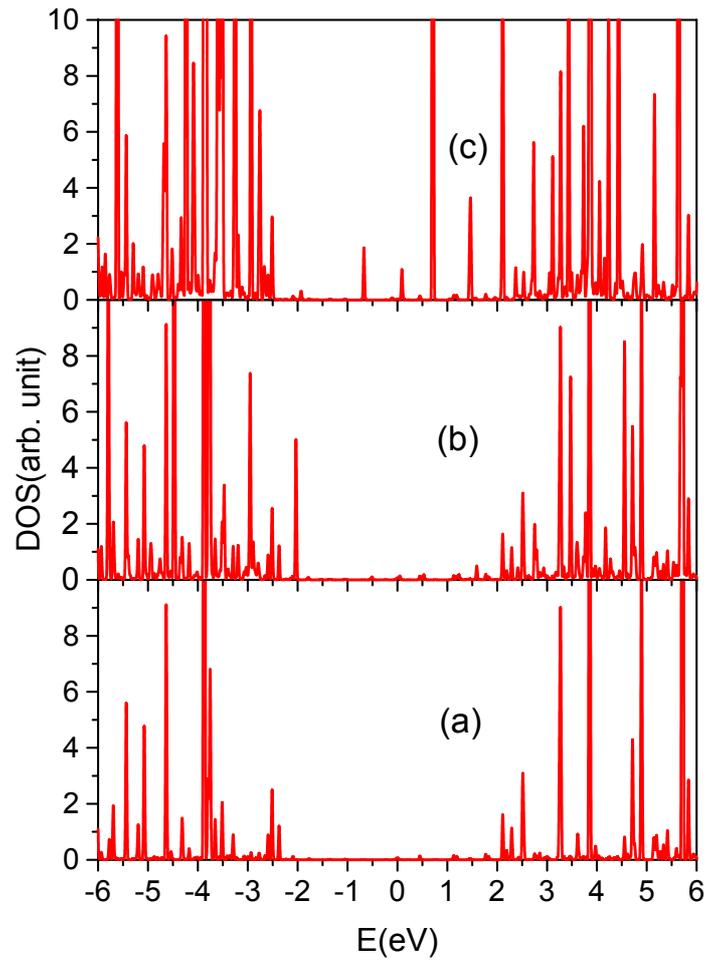

Fig. 2. Panels (a)-(c) show the electronic DOS vs. the Energy for the CNT/BCN/CNT structure for n=6, 12, 15 in (n,0) BCN nanotube and (n,0) CNTs, respectively. The length of BCN nanotube is selected nearly $5.92\ nm$.

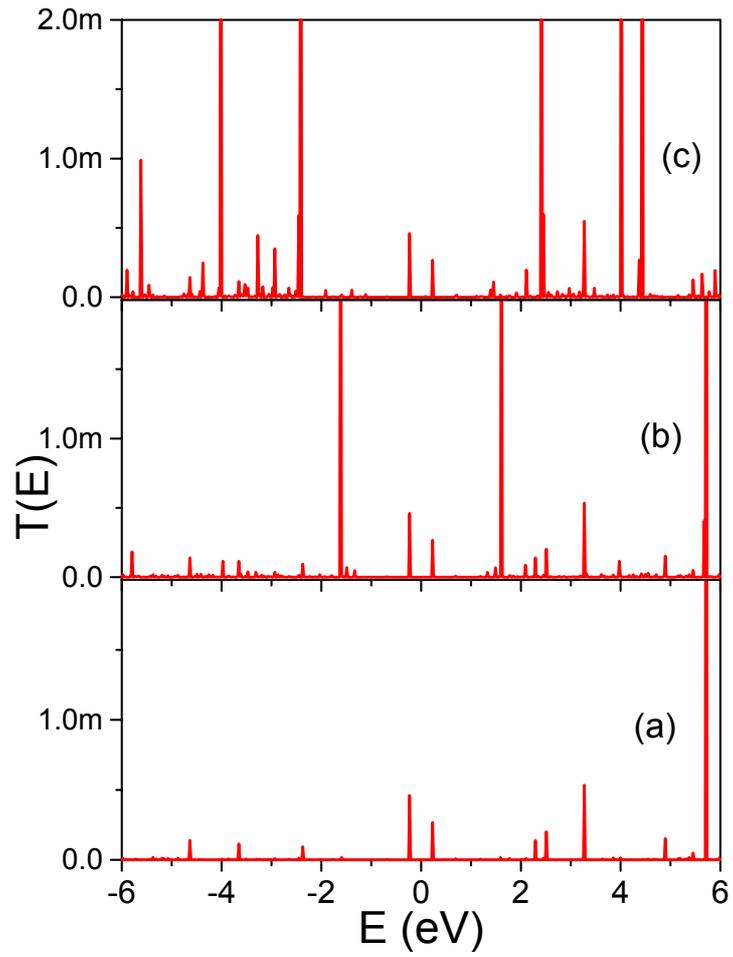

Fig. 3. Panels (a)-(c) show the transmission probability of an electron through the wire as a function of the Energy corresponding to plots (a)-(c) in Fig. 1, respectively.

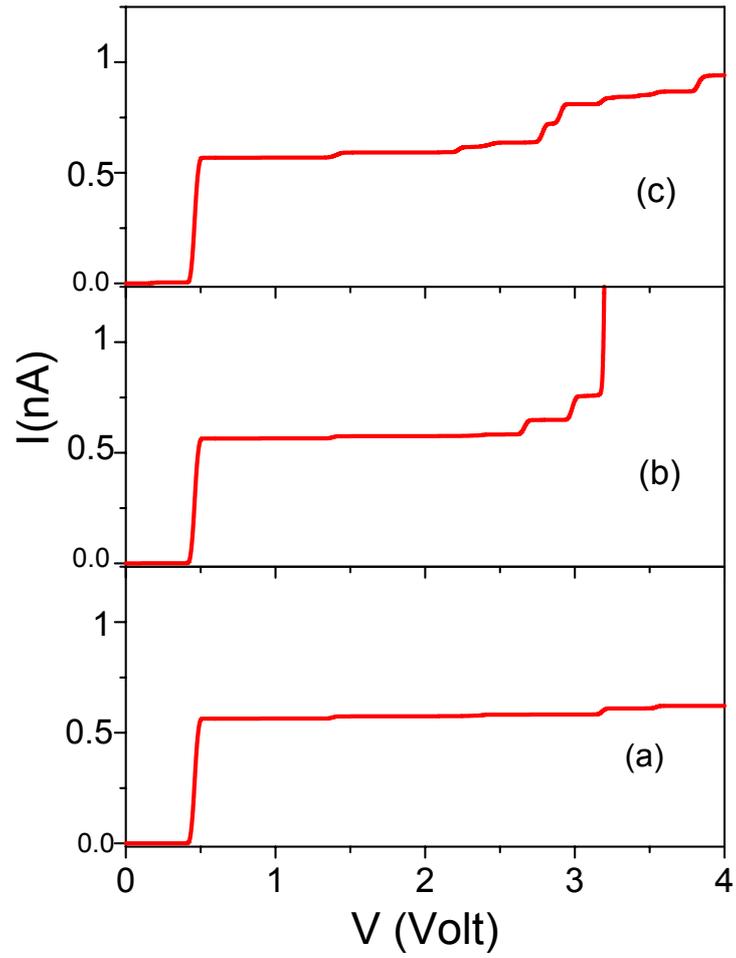

Fig.4. Panels (a)-(c) show the current-voltage characteristics of CNT/BCN/CNT structure corresponding to plots (a)-(c) in Fig. 1, respectively.

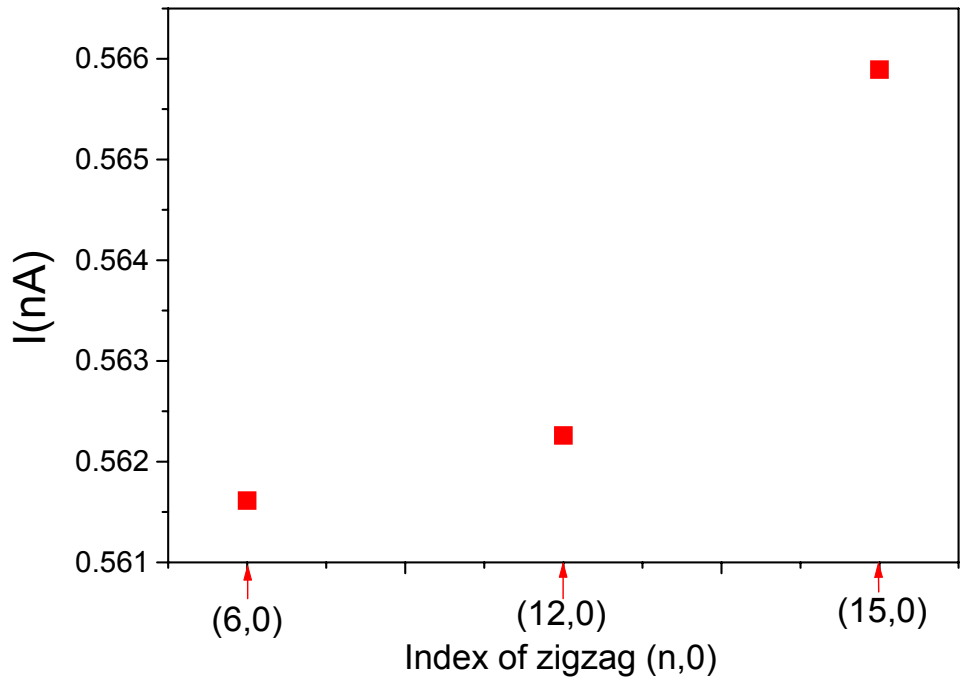

Fig. 5. The current versus zigzag nanotubes' diameter in V= 0.4985 (Volt).

**References**
[1] C.Kergueris and *et al.*, *Phys. Rev. B* **59** (1999) 12505.
[2] M.Magoga and C.Joachim, *Phys. Rev. B*  **56** (1998) 4722 and **59** (1999) 16011.
[3] E.G.Emberly and G.Kirczenow, *Phys. Rev. B* **58** (1998) 10911; *Phys. Rev. Lett.* **81** (1998) 5205; *Nanotechnology* **10** (1999) 285 ;  *J. Phys.: Condens. Matter* **11** (1999) 6911.
[4] L.E.Hall and *et al.*, *J.Chem. Phys.* **112** (2000) 1510.
[5] S.Datta and *et al.*, *Phys. Rev. Lett.* **79** (1997) 2530; W.Tian, *Physica E* (Amsterdam) **1** (1997) 304; *J.Chem.Phys.* **109** (1998) 2874; Y.Xue, *Phys. Rev. B* **59** (1999) R7852.
[6] Qi.Pengfei and *et al.*, *J.Am.Chem. Soc.* **126** (2004) 11774.
[7] X.Guo and *et al.*, *Nature Nanotechnology* **3** (2008) 163.
[8] R.Gutierrez and *et al.*, *Europhys. Lett.* **62** (2003) 90.
[9] Miriam del Valle and *et al.*, *Nature Nanotechnology* **2** (2007) 176.
[10] G.C.Liang and *et al.*, *Phys. Rev. B* **69** (2004) 115302.
[11] J.Guo and *et al.*, *IEEE Trans. Nanotech.* **2** (2003) 329.
[12] Kikuo Harigaya, *Physica E* **29** (2005) 628.
[13]  R.Saito, G.Dresselhaus, M.S.Dresselhaus, Physical Properties of Carbon Nanotubes, Imperial College Press, London, 1998.
[14] J.Otsuka, K.Hirose and T.Ono, *Phys. Rev. B* **78** (2008) 035426.
[15] C.T.White, D.H.Robertson, J.W.Mintmire, *Phys. Rev. B* **47** (1993) 5485.
[16] J.W.G.Wildoer and *et al.*, *Nature* (London) **391** (1998) 59.
[17] J.Yu, *Chem. Phys. Lett.* **323** (2000) 529 and therein.
[18] G.Y.Guo and *et al.*, *Phys. Rev. B* **69** (2004) 205416.
[19] F.Piazza and *et al.*, *Diamond and Related Materials* **14** (2005) 965.
[20] O.Stephan and *et al.*, *Science* **266** (1994) 1683.
[21] Z.Weng-Sieh and *et al.*, *Phys. Rev. B* **51** (1995) 11229.
[22] P.Redlich and *et al.*, *Chem. Phys. Lett.* **260** (1996) 465.
[23] K.Suenaga and *et al.*, *Science* **278** (1997) 653.

[24] M.Terrones and *et al*., *Chem. Phys. Lett.* **257** (1996) 576.
[25] R.Sen and *et al*., *Chem. Phys. Lett.* **287** (1998) 671.
[26] Y.Zhang and *et al*., *Chem. Phys. Lett*. **279** (1997) 264.
[27] M.Buttiker and *et al*., *Phys. Rev. B*  **31** (1985) 6207.
[28] V.Mujica ,M.Kemp and M.A.Ratner, *J. Chem. Phys.* **101** (1994) 6849.
[29] B.Larade and A.M.Bratkovsky, *Phys. Rev. B* **68** (2003) 235305.
[30] Y.Calev and *et al*., *Israel Journal of Chemistry* **44** (2004) 133.
[31] P.A.Khomyakov and *et al.*, *Phys. Rev. B* **72** (2005) 035450.
[32] W.Fa, J.Chen, H.Liu and J.Dong, *Phys. Rev. B* **69** (2004) 235413.
[33] X.Yang and J.Dong, *Phys. Lett. A* **330** (2004) 238.
[34] J.W.G.Wildoer and *et al.,* *Nature* **391** (1998) 59.
[35] P.Kim, T.W.Odom, J.L.Huang and C.M.Lieber, *Phys. Rev. Lett*. **82** (1999) 1225.